# On the low-field insulator-quantum Hall conductor transitions

Tsai-Yu Huang[1], J. R. Juang[1], C. F. Huang[2], Gil-Ho Kim[3,4], Chao-Ping Huang[1], C. -T. Liang[1], Y. H. Chang[1], Y. F. Chen[1], Y. Lee[3], and D. A. Ritchie[4]

[1]*Department of Physics, National Taiwan University, Taipei 106, Taiwan*

[2]*National Measurement Laboratory, Centre for Measurement Standards, Industrial Technology Research Institute, Hsin-Chu 300, Taiwan*

[3]*Department of Electronic and Electrical Engineering, Sungkyunkwan University, Suwon 440-760, Korea*

[4]*Cavendish Laboratory, Maglingley Road, Cambridge CB3 0HE, United Kingdom*

We studied the insulator-quantum Hall conductor transition which separates the low-field insulator from the quantum Hall state of the filling factor $\nu=4$ on a gated two-dimensional GaAs electron system containing self-assembled InAs quantum dots. To enter the $\nu=4$ quantum Hall state directly from the low-field insulator, the two-dimensional system undergoes a crossover from the low-field localization to Landau quantization. The crossover, in fact, covers a wide range with respect to the magnetic field rather than only a small region near the critical point of the insulator-quantum Hall conductor transition.

Insulator-quantum Hall conductor (I-QH) transitions have attracted much attention recently [1-10]. These transitions occur when two-dimensional (2D) systems enter quantum Hall states from the insulating state. According to selection rules in the global phase diagram (GPD) suggested by Kivelson, Lee, and Zhang [1], in the integer quantum Hall effect (IQHE) such transitions are between the quantum Hall state of the filling factor $\nu=1$ and the insulating state. To enter any integer quantum Hall state from the insulating state, therefore, a 2D system must pass through $\nu=1$ quantum Hall state. However, I-QH transitions between $\nu \geq 3$ quantum Hall states and the insulating state are observed [2-5]. It is shown by Hanein *et al*. [11] that the low-field I-QH transitions separating the integer quantum Hall liquid from the low-field insulator, in fact, can be linked to the 2D metal-insulator transition [12], which occurs at a zero magnetic field and is also inconsistent with the GPD.

For convenience, denotes the I-QH transition between the insulating state and the quantum Hall state of the filling factor $\nu$ as 0-$\nu$ transition [1,3,5] (Usually the insulating state is denoted by the number "0".). Song *et al*. [2] claimed that the low-field 0-$\nu$ transition with $\nu \geq 3$ are *phase* transitions contradicting to the GPD, and the numerical studies [13] show that such transitions can be due to that extended states are destroyed by the disorder at low fields. On the other hand, Huckestein [6] claimed that there is no contradiction and the low-field 0-$\nu$ transitions with $\nu \geq 3$ are only crossovers from weak localization to Landau quantization rather than *phase* transitions. Huckestein argued that under finite temperatures and/or finite sizes, Landau quantization is important if $B>1/\mu$ and hence from the Drude model the crossover should occur when

$$\rho_{xy}/\rho_{xx} (\sim \mu B) \sim 1, \quad (1)$$

where $\mu$ is the mobility. Such arguments can explain why Eq. (1) holds at the critical point of the low-field 0-$\nu$ transitions with $\nu \geq 3$ [2,6]. However, Huang *et al*. [5] and Sheng *et al*. [7] showed that such low-field I-QH transitions can have properties of phase transitions.

To further study the low-field I-QH transition inconsistent with the GPD, we performed a magneto-transport study on the gated 2D GaAs electron system containing self-assembled InAs quantum dots. We identified a crossover from the low-field localization to Landau quantization when the 2D system enters $\nu=4$ quantum Hall state directly from the low-field insulator. The point at which $\rho_{xy}/\rho_{xx}\sim 1$, is within the crossover as expected. However, such a crossover covers a wide range with respect to the magnetic field rather than only a small region around the critical point of the 0-4 transition. In addition, in our study the critical point of the 0-4 transition *is not* the point at which $\rho_{xy}/\rho_{xx}\sim 1$.

Figure 1 shows the sample structure that was grown by molecular-beam epitaxy on a GaAs (100) substrate and consists of a 20 nm wide GaAs/Al$_{0.33}$Ga$_{0.67}$As quantum well that is modulation doped on one side using a 40 nm spacer layer. The growth of the GaAs quantum well was interrupted at its center, and the wafer was cooled from 580 $^0$C to 525 $^0$C. The shutter over the indium cell was opened for 80 sec, allowing growth of 2.15 monolayers of InAs capped by a 5nm GaAs layer, and self–assembled InAs quantum dots were formed. The alloy Au/Ni/Cr was deposited onto the surface to serve as the front-gate. In this study, we set the gate voltage $V_g$=-0.07 V. Magneto transport measurements were performed with a top-loading He$^3$ system at temperatures (*T*'s) ranging from 0.52 to1.6 K in a 15 T superconductor magnet. A phase sensitive four-terminal ac lock-in technique was used with a current of 10 nA. At low temperatures, the sample behaves as an insulator in the sense that the longitudinal resistivity $\rho_{xx}$ increases as the temperature *T* decreases when the magnetic field *B*=0. From the low-field Hall measurement and SdH oscillations, the carrier concentration $n$=1.08 × 10$^{11}$ cm$^{-2}$.

Figure 2 shows the curve $\rho_{xy}(B)$ at the temperature $T$=0.52 K and the curves of $\rho_{xx}(B)$ at $T$=0.52-1.60 K when the gate voltage $V_g$=-0.07 V. At low magnetic fields, $\rho_{xx}$ increases as $T$ decreases and the 2DES behaves as an insulator. With increasing $B$, SdH oscillations [14] appear when $B > B_s$=0.48 T and $\rho_{xx}$ becomes $T$-independent at the magnetic field $B_c$=0.89 T. The $T$-dependences, in fact, are different on the both sides of $B_c$, and quantum Hall plateaus corresponding to $\rho_{xy}=h/2e^2$ and $h/4e^2$ are observed when $B > B_c$. Therefore, $B_c$ is the critical magnetic field of the I-QH transition to separate the low-field insulator from the quantum Hall liquid, and we can identify $\nu$=4 and 2 quantum Hall states from the corresponding Hall plateaus. [1] In the observed I-QH transition, the 2DES enters the $\nu$=4 quantum Hall state directly from the low-field insulator and hence such a transition is a low-field 0-4 transition, which is inconsistent with the GPD.

In Fig. 2, at higher $B$ the 2DES exhibits features of Landau quantization, including both the SdH oscillations and quantum Hall effect while at lower $B$ it behaves as an insulator due to the low-field localization. Since SdH oscillations and the low-field insulator can be identified when $B > B_s$=0.48 T and $B > B_c$=0.89 T, respectively, the region where $B_s < B < B_c$ correspond to the crossover from low-field localization to Landau quantization. The observations of SdH oscillations in the low-field insulator have also been reported by Smorchkova *et al*. [15] and Kim *et al*. [16]. Because we also observed the low-field 0-4 transition, we can examine how the 2DES enters quantum Hall state of $\nu$≥3 directly from the low-field insulator in such a crossover. The inset in Fig. 2 shows the curves of $\rho_{xx}$ and $\rho_{xy}$ when $B_s < B < B_c$. We can see that the magnetic field $B_a$, at which Eq. (1) holds, is in the crossover between the magnetic fields $B_s$ and $B_c$ and hence this crossover do occur when $\mu B \sim \rho_{xy}/\rho_{xx} \sim 1$ as argued by Huckestein [6]. However, the critical magnetic field $B_c$ of the 0-4 transition does not correspond to $B_a$, and the crossover region covers 0.41 T in $B$ rather only a small region near $B_a$ (or $B_c$). From our study, therefore, a 2D system undergoes a crossover from low-field localization to Landau quantization when it enters a quantum Hall state of $\nu \geq 3$ directly from the low-field insulator. Such a crossover, however, can cover a wide range in $B$ rather than a small region near the critical point. At the critical field $B_c$, in fact, in our study the ratio $\rho_{xy}/\rho_{xx}$ is about 1.5 and is larger than 1. We note that as reported by Hilke *et al.* [17] the criterion $\rho_{xy}/\rho_{xx} \sim 1$ does not hold at the critical point.

In conclusion, we observed a low-field insulator-quantum Hall conductor transition inconsistent with the GPD in the two-dimensional GaAs electron system containing self-assembled InAs quantum dots. To enter a quantum Hall state of $\nu \geq 3$ directly from the low-field insulator, in our study the two-dimensional system undergoes a crossover from the low-field localization to Landau quantization. The point at which $\rho_{xy}/\rho_{xx}=1$ is located within the crossover as expected. However, such a crossover can cover a wide range with respect to the magnetic field rather than only a small region around the critical point of the I-QH transition. In addition, the point at which $\rho_{xy}/\rho_{xx} \sim 1$ is not the critical point of the I-QH transition.

This work was funded by the NSC, Taiwan, the MOE program for Promoting Academic Excellence of Universities (89-N-FA01-2-4-3), and in part, by the KOSEF through the Quantum Photonic Science Research Centre at Hanyang University. C. T. L. acknowledges financial support from the Department of Physics, National Taiwan University. G. H. K. acknowledges support by National R&D Project for Nano Science and Technology (Contract No. M1-0212-04-0003) of MOST.

Figure Captions

Fig. 1. The structure of the sample.

Fig. 2. The curves of $\rho_{xx}$ (B) at T = 0.52 – 1.60 K. The curve $\rho_{xy}$ (B) at T=0.52 K. The inset shows the curves between the magnetic $B_s$ and $B_c$.

Fig. 1

Tsai-Yu Huang *et al*.

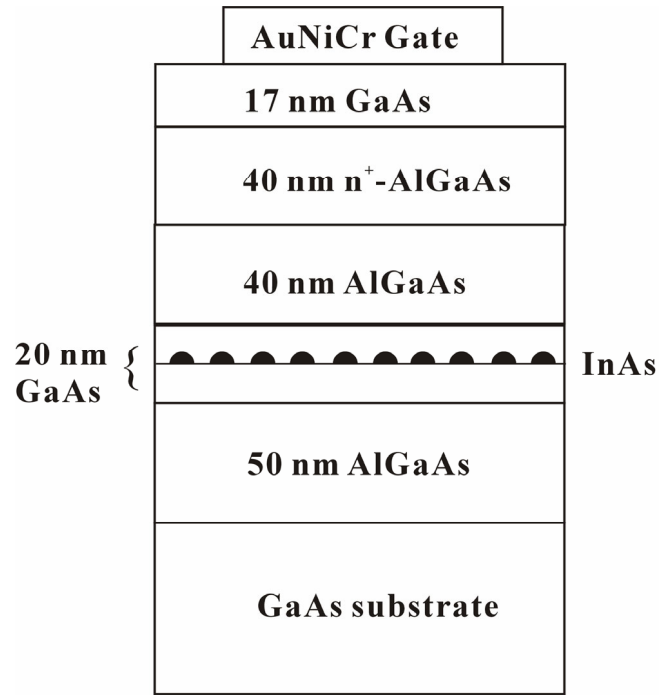

Tsai-Yu Huang *et al*.

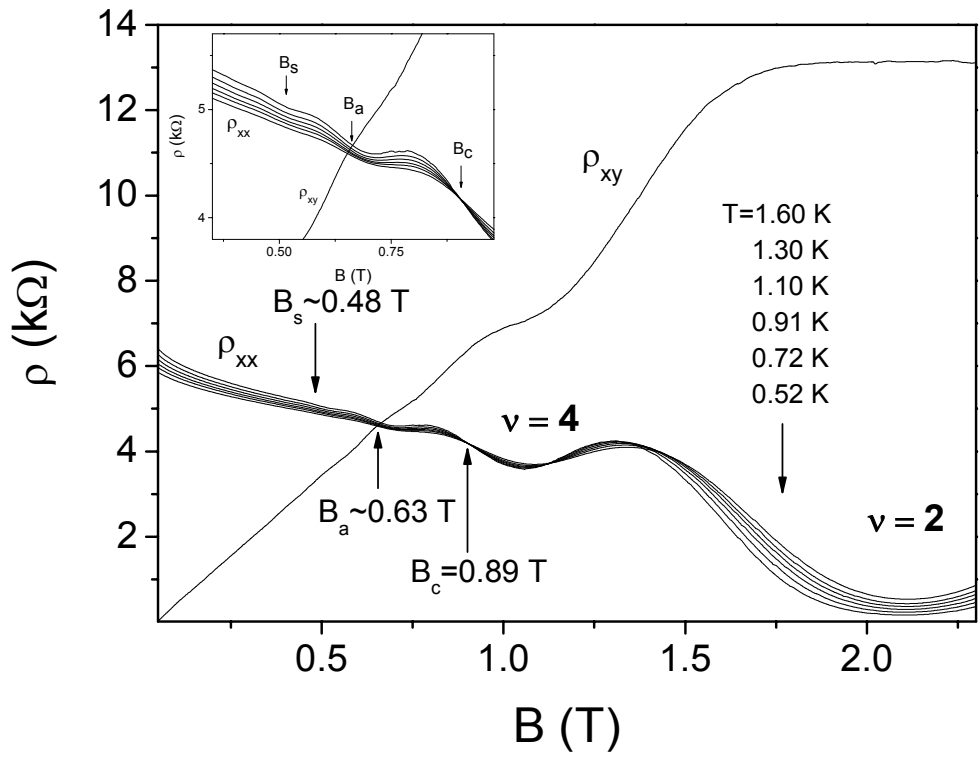